# Electronic structure and molecular orientation of a Zn-tetra-phenyl porphyrin multilayer on Si(111)

C. Castellarin Cudia[1], P. Vilmercati[1], R. Larciprete[1,2], C. Cepek[3], G. Zampieri[1,4], L. Sangaletti[5], S. Pagliara[5], A. Verdini[3], A. Cossaro[3], L. Floreano[3], A. Morgante[3], L. Petaccia[1], S. Lizzit[1], C. Battocchio[6], G. Polzonetti[7] and A. Goldoni[1]

1) Sincrotrone Trieste S.C.p.A, s.s. 14 Km 163.5 in Area Science Park, 34012 – Trieste, Italy

2) CNR-Istituto dei Sistemi Complessi, Via Salaria Km 29.3, 00016 Monterotondo (RM), Italy

3) Lab. Nazionale TASC-INFM, s.s. 14 Km 163.5 in Area Science Park, 34012 – Trieste, Italy

4) Laboratorio de Superficies, Centro Atómico Bariloche, 8400 S.C. de Bariloche, Argentina

5) Dipartimento di Matematica e Fisica, Universita` Cattolica del Sacro Cuore, Via dei Musei 41, 25121 Brescia, Italy

6) INFM-Ogg, c/o ESRF, GILDA CRG, 6, Rue Jules Horowitz, F-38043 Grenoble

7) INFM-Dipartimento di Fisica, Universita` "Roma Tre", Via della Vasca Navale 84, 00146 - Rome, Italy

## ABSTRACT

The electronic properties and the molecular orientation of Zn-tetraphenyl-porphyrin films deposited on Si(111) have been investigated using synchrotron radiation. For the first time we have revealed and assigned the fine structures in the electronic spectra related to the HOMOs and LUMOs states. This is particularly important in order to understand the orbital interactions, the bond formation and the evolution of the electronic properties with oxidation or reduction of the porphyrins in supramolecular donor-acceptor complexes used in photovoltaic devices.

# Introduction

Porphyrins are a ubiquitous class of naturally occurring molecules involved in a wide variety of important biological processes ranging from oxygen transport to photosynthesis, from catalysis to pigmentation changes [1]. The common feature of all these molecules is the basic structure of the macrocycle, which consists of four pyrrolic subunits bridged by four (*meso*) carbon atoms (see Fig.1). This macrocycle is an aromatic system, the size of which is perfect to bind almost all metal ions and, indeed, a number of metals (e.g. Fe, Zn, Cu, Ni, and Co) can be inserted in the center of the macrocycle forming metallo-porphyrins.

Porphyrin-based fundamental biological representatives include hemes, chlorophylls, vitamin B-12, and several others. Heme proteins (which contain iron porphyrins) serve many roles, like $O_2$ storage and transport (myoglobin and hemoglobin), electron transport (cytochromes b and c), and $O_2$ activation and utilization (cytochrome P450 and cytochrome oxidase). Chlorophylls (which have a central magnesium ion) and pheophytins (which are metal free) are found in the photosynthetic apparatus of plants and bacteria, while vitamin B-12 (which has cobalt) is catalytically active in bacteria and animals.

Given the capabilities of porphyrins to bind and release gases and to act as active center in catalytic reactions in biological systems, porphyrin-based films on surfaces are extremely appealing as chemical and gas sensors [2] as well as nanoporous catalytic materials [3] in novel synthetic bio-mimetic devices. Moreover, the role of porphyrins in photosynthetic mechanisms indicates a good attitude of these molecules to mediate visible photon – electron energy transfer processes [4]. For this reason, in recent years, metallo-porphyrins- and porphyrin-substrates (in particular Si, $TiO_2$ and noble metals) interfaces have become of major interest for applications in opto-electronics, data storage and solar cells [5] and a still increasing number of covalently linked donor-acceptor supramolecular porphyrin-based assemblies have been studied for these purposes [6].

Despite the interest and the huge amount of works on this family of molecules, there are only very few experimental articles on the growth *via* molecular beam epitaxy of thin films on various substrates and the investigation of the electronic properties using a direct probe like photoemission spectroscopy [8-10]. This paper focuses on an experimental study of the growth and electronic properties of a prototype molecule of this family, namely Zn-tetra-phenyl-porphyrin (ZnTPP), in the form of multilayer films on Si(111).

**Experimental**

The measurements were performed in the ultra-high-vacuum (UHV) experimental chambers - base pressure $10^{-10}$ mbar - of the SuperESCA and ALOISA beamlines at the Elettra Synchrotron facility. Highly purified ZnTPP (99,999%) were sublimated in UHV and deposited on the clean Si(111)-7x7 substrate kept at room temperature using a resistively heated Ta evaporator. The evaporation rate was about one monolayer (1 ML) every 10 min. Absorption geometry, growth and electronic properties were investigated by means of valence band and core level photoemission and x-ray absorption spectroscopy (XAS). Photoemission spectra were obtained by collecting the photoelectrons with a double-pass hemispherical analyzer having an angular acceptance of ±1˚, while XAS spectra were measured at the C 1s and N 1s thresholds in partial yield by collecting the C and N Auger electrons, respectively. The energy resolution was 50 meV for valence band photoemission and XAS, while it was ~100 meV for core level photoemission.

**Results and discussion**

ZnTPP is known to form a monoclinic molecular crystal, with an effectively flat macrocycle ($D_{4h}$ point group) [11]. The $D_{4h}$ symmetry of the molecule in the film is confirmed also by the N 1s photoemission spectrum (not reported here) showing a single peak with a full-width at half-maximum of 0.7 eV, which indicates that the four N atoms are equivalent. This is at variance, for example, with the N 1s core level spectra of $H_2TPP$ [10] (i.e. without metal), where two components separated by about 2 eV are observed.

Because of the steric interaction between the hydrogen atoms, the plane of the four phenyl groups forms an angle between 70° and 90° degrees with respect to the macrocycle, minimizing in such a way the interaction between the phenyl and macrocycle π systems [4]. The presence of the phenyl groups, therefore, should cause only a very weak perturbation of the macrocycle states, and the electronic structure is expected as a rough superposition of the two electronic state systems (macrocycle and phenyl groups). From this point of view, it is interesting to look at the C 1s photoemission and XAS spectra. Fig. 2a compares the XAS spectrum at the C 1s threshold of 20 ML of ZnTPP/Si(111), measured with the linear polarization of the light at 20° from the surface normal, with the corresponding XAS spectra of benzene [12] (dashed line) and Zn-octaethyl-porphyrin (i.e. no phenyl rings, with dominant contribution from the macrocycle, dotted line) [12]. The ZnTPP spectrum looks like a simple superposition of the other two, confirming that there is almost no interaction between the π states of the macrocycle and those of the phenyl groups. The near edge region is shown in the inset of Fig. 2a where the above ZnTPP XAS spectrum is compared to the one taken at normal incidence (i.e. with the light polarization on the surface plane). The first absorption peak reflects the electronic transitions from the C 1s core level to the lowest unoccupied molecular orbital (LUMO), which mainly belongs to the macrocycle π* system. The LUMO peak has $e_g$ symmetry and is, therefore, two-fold degenerate [13]. We can make two main observations: i) There is a clear polarization dependence of the LUMO intensity, indicating that the molecules are oriented; ii) The LUMO is made by two features separated by 0.25 eV.

We will discuss the molecular orientation later, while here we address the origin of the two LUMO features. One possible explanation is the splitting of the two degenerate LUMO states due to Jahn-Teller distortion and/or ZnTPP conjugation in the lattice structure, but this splitting should be of the order of a few tens of meV only. The only remaining possibility is the presence of different C 1s initial states, bringing us to the analysis of the C 1s core level. Fig. 2b shows the C 1s core level photoemission spectrum of ZnTPP as measured at normal emission with photon energy of 400 eV. The full width at half maximum (~ 1 eV) and the lineshape (with an anomalous asymmetry on the low binding energy side) reveal the presence of several components. There are 24 carbon atoms belonging to the phenyl groups and 20 carbon atoms belonging to the macrocycle. The latter can be separated in *meso*- carbon bridges (4 atoms) and pyrrolic carbon atoms (16 atoms, 8 of which having a C-N bond). By fitting the core level spectrum with four components having the same width and fixed intensity ratios 24:8:8:4, we obtain a good fit as shown in Fig. 2b and the following binding energy positions: 284.80 eV (phenyl), 285.08 eV (C-C-N, pyrrolic), 284.27 (C-C-C, pyrrolic) and 284.02 eV (*meso*-bridges). The energy separation between the last two components is exactly 0.25 eV, confirming that the fine structure of the LUMO peak in the XAS spectrum is due to the presence of inequivalent C atoms in the macrocycle. Moreover, while the lowest transitions from the phenyl atoms are expected to mainly contribute to the main XAS peak, a further structure is expected due to the transition into the LUMO from the pyrrolic carbon atoms bonded to nitrogen. This should produce a feature in the XAS spectrum at ~ 1 eV from the lowest transition, which therefore is hidden under the main absorption peak of the phenyl atoms. Since the phenyl atom contribution to the XAS spectrum is not expected to have pronounced polarization dependence due to the presence of several orientations of the phenyl rings, by subtracting the two ZnTPP spectra reported in the inset we should almost cancel this contribution and enhance the macrocycle contribution. The difference spectrum is also reported in the inset (dashed line). As expected this spectrum looks like the XAS spectrum of the Zn-octaethyl-porphyrin and a feature at ~ 1 eV from the first transition is now visible, which should correspond to the C 1s-LUMO transition for the pyrrolic C atoms bonded to N.

Further information on the electronic structure is obtained from the analysis of the occupied valence states. Fig. 3a shows the valence band photoemission spectrum of the ZnTPP multilayer measured at hν=91 eV in normal incidence conditions (grazing emission), by collecting the electrons at 70° from the surface normal. Apart the intense feature at ~ 11 eV that is dominated by the Zn 3d emission, the valence band is mainly formed by peaks corresponding to π and σ states of the ZnTPP macrocycle. In the inset, the region near the Fermi level ($E_F$) is compared for two experimental geometries: normal incidence (linear polarization in the surface plane) by collecting the electrons at 70° from the surface normal, and grazing incidence (linear polarization at 70° from the surface plane) by collecting the photoelectrons at normal emission. There are two things that meet the eye. First, there is a clear dependence from the experimental geometry, that may be given either by band dispersion or by polarization dependence in the photoemission initial state. Second, the peak at ~ 2.5 eV, that was recognized as the highest occupied molecular orbital (HOMO) in other photoemission experiments [8], has a shoulder at lower binding energy, suggesting that this shoulder is the real HOMO emission.

To better understand, underneath the experimental spectra we report also the simulated occupied outer molecular orbital emissions (gaussian peaks, with σ = 0.275 eV), whose energy separations are assumed according to Liao and Scheiner [13] by putting the HOMO level at 1.25 eV below $E_F$, and whose intensity is proportional to the number of orbitals, including degeneracy, at each energy. The sum of these states, which simulates the total DOS uncorrected by the matrix effects, is also shown. The agreement with the experimental data is pretty good, in particular with the data collected at normal emission, confirming our assignment of the HOMO emission and that the main changes observed by changing the experimental geometry are due to polarization (matrix) effects in the initial state. These observations led to the energy level alignment shown in Fig. 3b by assuming the energy levels calculated by Liao and Scheiner [13] for the isolated free ZnTPP. We note that molecular conjugation and polarization screening in the molecular film may reduce the HOMO-LUMO gap and the ionization potential compared to the isolated free molecule, so the interface dipole Δ ~ 0.95 eV as schematized in Fig. 3b represents the lower limit.

Finally, we address the problem of the molecular orientation. The XAS spectra measured at the C 1s edge show a notable dependence from the incidence angle of the light, indicating that the molecular macrocycles in the film are oriented, maybe azimuthally disordered, but with a defined angle with respect to the surface plane. However, since the N atoms belong to the macrocycle only and their $p_z$ orbitals are perpendicular to the ZnTPP macrocycle plane, the measurement of XAS spectra at the N 1s edge is more useful to obtain the molecular orientation. We measured the full angular dependence of the N 1s XAS spectra by keeping the surface at a constant grazing incidence angle of 6° and by rotating the sample around the beam axis by steps of 10°, moving in such a way the light polarization from in-plane to almost normal to the surface plane. Fig. 4 shows these spectra for the ZnTPP multilayer, while the inset shows the area of the first π* absorption peak as a function of the angle θ between the polarization and the surface. The fact that the π* transitions peaks vanish when the polarization of the light lies in the surface plane is a strong evidence that the ZnTPP molecules grow with the macrocycle parallel to the Si(111) surface plane. Indeed, by fitting the observed behaviour to I(θ,Φ)=A($\sin^2\theta \cdot \sin^2\Phi + 0.5 \cdot \cos^2\theta \cdot \cos^2\Phi$) (valid for three or more azimuthally rotated domains) [14] - see figure for angle definitions - we obtain that the ZnTPP macrocycle forms an angle of 5°±5° with the sample surface (or equivalently, the π* orbitals form an angle Φ=85°±5°). The same result is obtained assuming a simple $\cos^2$ dependence (single domain) and the same kind of growth (i.e. the ZnTPP macrocycle parallel to the substrate plane) has been observed on Ag(110) and Si(100) [15].

**Conclusions**

Using synchrotron radiation photoemission and x-ray absorption spectroscopy we have investigated the electronic properties and the molecular orientation of ZnTPP films (more than 20 ML) deposited in UHV on Si(111). For the first time we have revealed, identified and assigned the fine structures in the electronic spectra related to the HOMOs and LUMOs states. This is particularly important in order to understand orbital interactions, bond formation and evolution of the electronic properties with doping (oxidation and reduction) in the light of possible applications of porphyrins in donor-acceptor complexes for photovoltaic devices or, given the high molecular symmetry, as prototypical systems for the verification of recent models and phase diagrams for strongly correlated materials [16].

# References

[1] *Porphyrins and Metalloporphyrins*, K. M. Smith, Ed., Elsevier, Amsterdam (1975); M. Boulton et al., J. Photochem. & Photobio. B: Biology **64**, 144 (2001); E.I. Sagun et al, Chem. Phys. **275**, 211 (2002).

[2] J.P. Collman et al., in *Metal Ions in Biology* **Vol. 2**, T.G. Spiro Ed., Wiley, NY, 1 (1980); M.A. Schiavon et al., J. Molecular Catalysis A **174**, 213 (2001).

[3] K.S. Suslick, et al., in *Comprehensive Supramolecular Chemistry*, **Vol. 5**; K.S. Suslick Ed., Elsevier, Oxford, 141 (1996); P. Bhyrappa et al., J. Am. Chem. Soc. **118**, 5708 (1996); J.A.A. Elemans et al., Org. Chem. **64**, 7009 (1999).

[4] H.L. Anderson, Chem. Commun., 2323 (1999).

[5] K. Yamamshita et al., J. Phys. C: Solid State Phys. **93**, 5311 (1989); J. Chen et al., Science **286,** 1550 (1999); A. Tsuda and A. Osuka, Science **293**, 79 (2001).

[6] D. Gust et al., Topics in Current Chemistry **159**, 103 (1991); W. Han et al., J. Phys. Chem. B **101**, 10719 (1997); G. Stainberg-Yfrach et al., Nature **392**, 479 (1998).

[7] H. Ishii et al., J. Electron Spectr. Related Phenom. **76**, 559 (1995); D. Yoshimura et al., J. Electron Spectr. Related Phenom. **78**, 359 (1996); K. Seki et al., Synth. Metals **91**, 137 (1997).

[8] S. Narioka et al., J. Phys. Chem. **99**, 1332 (1995).

[9] A. Ferri et al., Surf. Interface Anal. **30**, 407 (2000); G. Polzonetti et al., Chem. Phys. **296**, 87 (2004).

[10] Y. Niwa, H. Kabayashi, T. Tsuchya, J. Chem, Phys. **60**, 799 (1974).

[11] M.P. Byrn et al., J. Am. Chem. Soc. **115**, 9480 (1993).

[12] G. Polzonetti, private communication

[13] M.-S. Liao and S. Scheiner, J. Chem. Phys. **117**, 205 (2002).

[14] J. Stöhr, in *NEXAFS Spectroscopy*, Springer Series in Surface Science vol. **25**,

Springer, Berlin (1992).

**[15]** C. Castellarin-Cudia et al., unpublished

**[16]** M. Capone et al., Phys. Rev. Lett. **93**, 047001 (2004).

**FIGURE CAPTIONS**

**Fig. 1:** Schematic view of a ZnTPP molecule.

**Fig. 2: a)** XAS spectra at the C 1s threshold of ZnTPP (solid line), benzene (dashed line) and Zn-octaethyl-porphyrin (dotted line). **Inset:** Near-edge region of the ZnTPP XAS spectra collected at normal incidence and at grazing incidence (20° from the surface plane) of the linearly polarized light. The difference spectrum is also shown (dashed line). The arrows indicate the XAS transitions into the LUMO due to the inequivalent C atoms of the macrocycle (*meso*-bridges, pyrrol C-C-C and pyrrol C-C-N, from left to right, respectively). **b)** C 1s photoemission spectrum of ZnTPP (dots) with the corresponding best fit (solid line). The binding energy and intensity of the four components used in the fit are shown as bars (see text for details).

**Fig. 3: a)** Valence band photoemission spectrum of ZnTPP ($E_F$=0 eV). **Inset:** The ZnTPP photoemission spectra near $E_F$ collected at normal emission (grazing incidence, dashed line) and normal incidence. Under the experimental spectra we report the expected contributions of HOMO, HOMO-1, HOMO-2, HOMO-3 and HOMO-4 states (from bottom to top) and the corresponding sum (DOS). The energy separation between these states has been taken in agreement with ref. [13]. **b)** Schematic energy level diagram for the ZnTPP multilayer on Si(111) in agreement with our experimental data and assuming the energy levels of the isolated ZnTPP molecule in agreement with ref. [13].

**Fig. 4:** Angular dependence of the ZnTPP XAS spectra at the N 1s threshold. A scheme of the experimental geometry is shown. **Inset**: Area of the N 1s→LUMO transition as a function of the angle θ (corrected for a tilt angle of 6°) between the polarization vector and the sample surface. The plot is normalized to the area at θ = 90°.

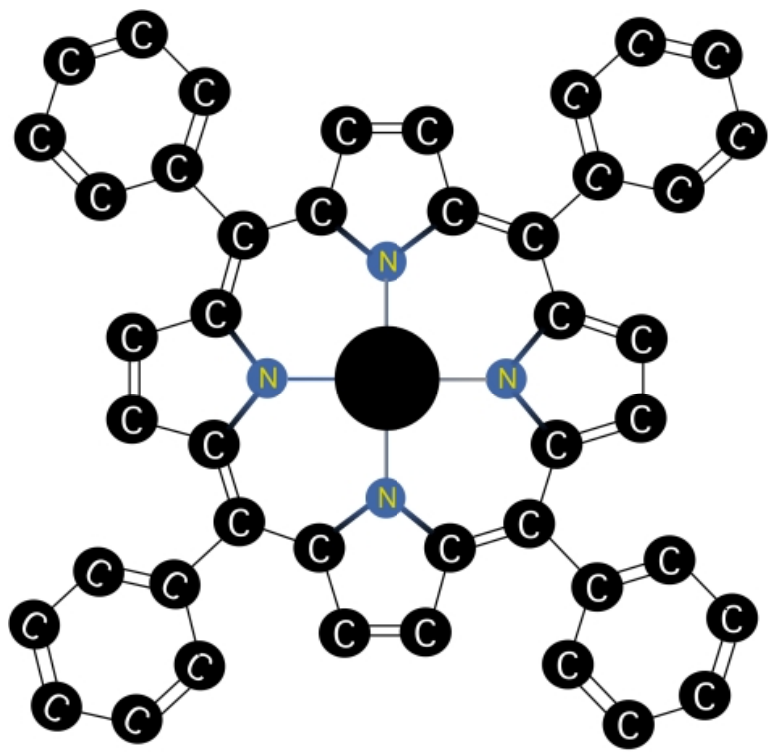

C. Castellarin Cudia et al., Figure 1

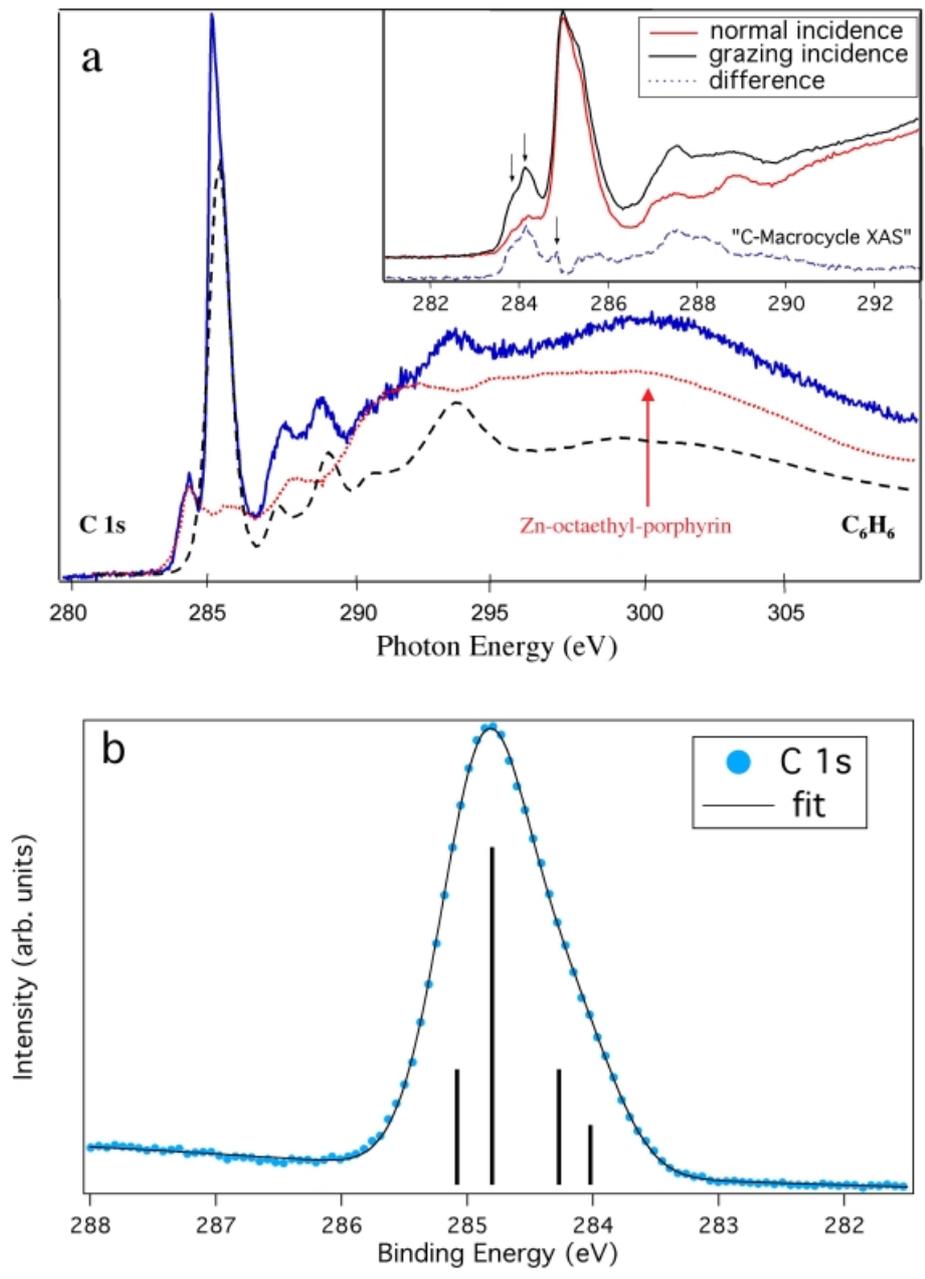

C. Castellarin Cudia et al., Figure 2

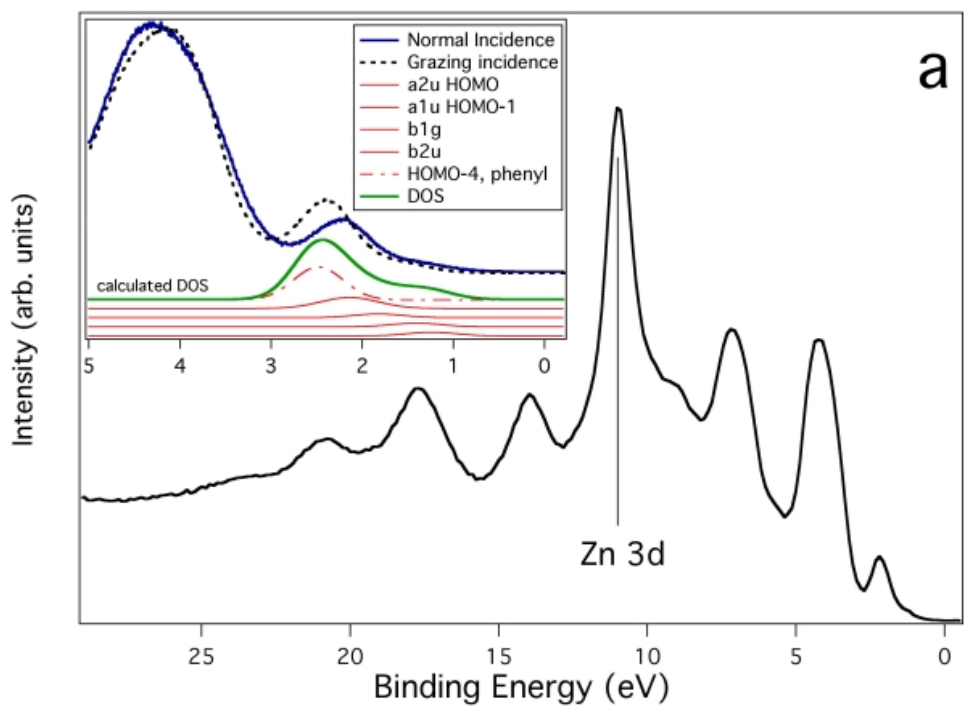
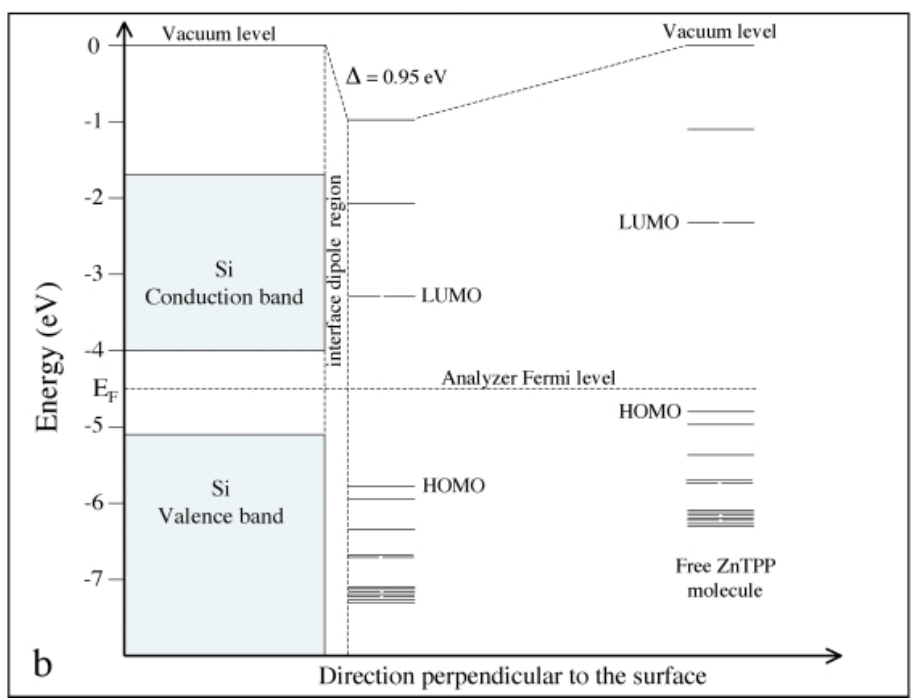

C. Castellarin Cudia et al., Figure 3

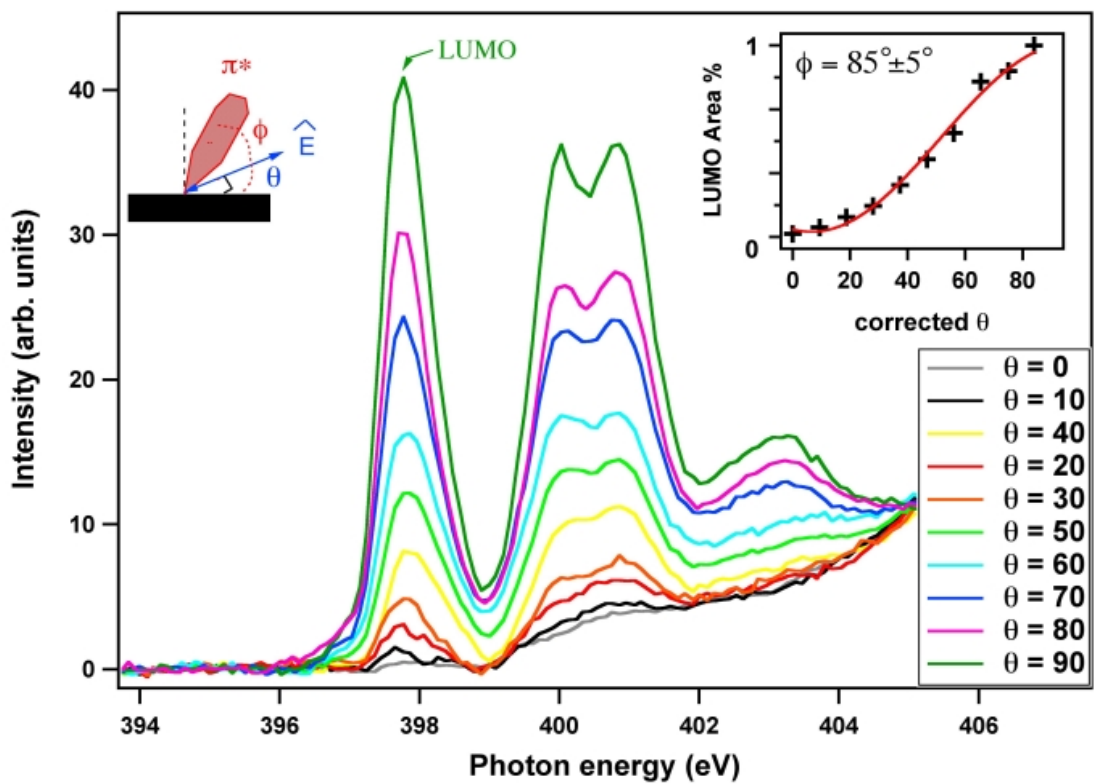

C. Castellarin Cudia et al., Figure 4